\documentclass[final,5p,times,twocolumn]{elsarticle}

\usepackage{amssymb}
\usepackage{amsthm}

\usepackage{amsmath}
\usepackage{tabularx}
\newcolumntype{C}[1]{>{\centering\arraybackslash}p{#1}}
\usepackage{dcolumn}
\usepackage{siunitx}
\usepackage{color}
\usepackage[capitalise]{cleveref}
\usepackage{tikz}
\usepackage{pgfplots}
\usepgfplotslibrary{groupplots}
\usetikzlibrary{spy,decorations.fractals,shapes.misc}
\usetikzlibrary{decorations.markings}
\usepgfplotslibrary{colormaps}
\usepgfplotslibrary{colorbrewer}
\pgfplotsset{cycle list/Dark2}

\usepackage[caption=false]{subfig}
\usepackage{epstopdf}
\usepackage{booktabs,multirow}
\usepackage{xcolor}

\journal{Journal of Computational Physics}

\begin{document}

\begin{frontmatter}



\title{Multi-species modeling in the particle-based ellipsoidal statistical Bhatnagar-Gross-Krook method including internal degrees of freedom}


\author[a]{F. Hild}
\author[a]{M. Pfeiffer}

\affiliation[a]{organization={Institute of Space Systems, University of Stuttgart},
            addressline={Pfaffenwaldring 29}, 
            city={Stuttgart},
            postcode={70569},
            country={Germany}}

\begin{abstract}
The implementation of the ellipsoidal statistical Bhatnagar-Gross-Krook (ESBGK) method in the open-source particle code PICLas is extended for multi-species modeling of polyatomic molecules, including internal energies with multiple vibrational degrees of freedom.
For this, the models of \citet{MATHIAUD202265,energy-cons} and \citet{brull,brull2021ellipsoidal} are combined.
In order to determine the transport coefficients of the gas mixture, Wilke's mixing rules and collision integrals are compared.
The implementation is verified with simulation test cases of a supersonic Couette flow as well as a hypersonic flow around a 70° blunted cone.
The solutions of the ESBGK method are compared to the Direct Simulation Monte Carlo (DSMC) method to assess the accuracy, where overall good agreement is achieved.
In general, collision integrals should be preferred for the determination of the transport coefficients, since the results using Wilke's mixing rules show larger deviations in particular for large mass ratios. 
Considering the computational efficiency of the methods, a considerable reduction in computational time is possible with the ESBGK model.
\end{abstract}




\end{frontmatter}


\section{Introduction}
Numerical simulations of fluid dynamics in space applications, micro and nano flows, as well as vacuum technology present significant challenges for applied numerical methods.
These applications involve large density gradients, ranging from the continuum to the free molecular flow regime, and require versatile numerical approaches due to their multi-scale and non-equilibrium nature.
Although a highly accurate solution of these flows can be achieved by using the well-established Direct Simulation Monte Carlo (DSMC)~\cite{bird} method, the latter requires excessive computational effort in the transition and continuum regimes.
Thus, the coupling of DSMC with a computationally efficient method is desirable.

For continuum flows, computational fluid dynamics (CFD) are typically used.
However, many problems arise in the coupling with the DSMC method due to the very different underlying approaches of both methods.
Especially the statistical noise of the DSMC method is problematic at the boundaries between DSMC and CFD~\cite{zhang2019particle,burt-boyd-ld,pfeiffer2019evaluation,SCHWARTZENTRUBER2006402}.

Particle-based continuum methods have emerged as promising alternatives, bridging the transition regime.
Recent developments include but are not limited
to the Fokker-Planck approach~\cite{hossein,fp,mathiaud2016fokker,Jun2019} and the Bhatnagar-Gross-Krook (BGK)~\cite{bgk,gallis-torczynski-2,zhang2019particle,burt2006evaluation} model, which is the focus of this paper.
The latter approximates the collision integral of the Boltzmann equation by a relaxation process.
With this, the mean free path and the collision frequency do not need to be resolved, which means that the choice of time step and particle weighting factor for particle simulations can be less restrictive.
The necessary local resolution is rather defined by the gradients of the macroscopic moments of the distribution function present in the flow, such as density, velocity or temperature, similar to that described in~\citet{fp}.
The time step is then determined by the stiffness of the BGK collision term, i.e. the relaxation frequency to be resolved.
The advantage over DSMC, however, is that the BGK equation no longer represents a jump process but can use different time integration and space interpolation methods to achieve significantly coarser resolutions.
While this has been a focus of research in connection with BGK in the context of discrete velocity methods~\cite{guo2013discrete,mieussens2000discrete} for a long time, progress has also been made in recent years on the particle methods side~\cite{PhysRevE.106.025303, fei2021efficient, FEI2020108972, liu2020unified}.
Thus, less computational costs are expected compared to DSMC simulations, in particular for low-$Kn$ regimes.

The particle-based ellipsoidal statistical BGK (ESBGK)~\cite{esbgk} and the Shakhov BGK~\cite{shakhov} methods were investigated in particular~\cite{piclas-bgk,fei2020benchmark}, both producing the correct Prandtl number of the gas.
It was showed that in some cases the ESBGK model is more robust and more efficient in the particle context~\cite{piclas-bgk, pfeiffer2019evaluation}.
In addition, the ESBGK method fulfills the H-theorem~\cite{andries2000gaussian}, which is why this approach is followed here.

An overview of different BGK models for gas mixtures is available from~\citet{overview-bgk-mix}.
A rough distinction is made between two types of mixture models.
The first type are models with a sum of several relaxation terms, so that each species combination has its own relaxation term~\cite{asinari2008asymptotic,garzo1989kinetic,hamel1965kinetic,klingenberg2018consistent,klingenberg2018kinetic,bobylev2018general}.
The advantage here is that the correct collision rates and thus energy/moment exchange rates between the individual species can theoretically be reproduced correctly according to the Boltzmann equation.
The disadvantage, however, is that the models are significantly more complex and therefore require more computing time.
In addition, the choice of free parameters such as the different relaxation frequencies is challenging from a physical point of view, for example if certain viscosities or Prandtl numbers are to be achieved for the entire mixture.
This is especially true for multi-species mixtures with more than two components and when additional exchange terms for internal degrees of freedom are added in the polyatomic case.
On the other hand, there are models with only one relaxation term on the right hand side~\cite{andries2002consistent,brull,brull2021ellipsoidal,todorova-shakhov-esbgk,todorova2020}.
The obvious disadvantage is that the correct energy and momentum exchange times cannot be modeled with this approach.
However, there are some advantages for the practical application of the models.
The models are less complex and therefore require less computing time.
In addition, the collective behavior of the gas mixture, such as viscosity, heat flow, Prandtl number, etc., can be modeled correctly with significantly fewer parameters, whereby established mixing rules can be used.
Since the left-hand side in the form of advection is still modeled correctly, the error due to the use of only one relaxation term can be small, depending on the application.

The focus of this paper is the extension of the particle-based ESBGK model to gas mixtures including non-equilibrium states of internal degrees of freedom of di- and polyatomic molecules, including quantized vibrational states by using a model with only one relaxation term.
The work is based on the models of \citet{MATHIAUD202265,energy-cons} and \citet{brull,brull2021ellipsoidal}. 
In the applications presented here, this model using only one relaxation term is already capable of accurately representing non-equilibrium effects in complex 2D flows.
This was already demonstrated in~\citet{bgk-multispecies} using pure atomic mixtures and has been successfully extended here to mixtures with any number of atomic and polyatomic species, including non-equilibrium effects of internal excitation energies.
In the models with one relaxation term mentioned above, either binary molecular mixtures~\cite{todorova2020}, atomic mixtures~\cite{andries2002consistent,todorova-shakhov-esbgk,brull}, or equilibrium in the internal excitation states~\cite{brull2021ellipsoidal} have been assumed so far.
In the specific case of stochastic particle methods, models with only one relaxation term have the additional advantage that solely the moments of the entire mixture, not the individual species, are required.
Thus, a significantly smaller particle representation of the distribution function, which means fewer simulation particles, can be used compared to the case where all individual moments of the distribution function are needed, as it is in the case of multi-relaxation term models.

\section{Theory}
The BGK operator is an approximation of the Boltzmann collision integral as relaxation process of the particle distribution function $f_s \left(\mathbf x,\mathbf v, t \right)$ of a species $s$ at position $\mathbf x$ and with velocity $\mathbf v$ towards a target distribution $f_s^t$~\cite{bgk}
\begin{equation}
	\partial_t f_s + \mathbf {v} \partial_{\mathbf x} f_s = \nu(f_s^t-f_s), \label{eq:BGK}
\end{equation}
using the relaxation frequency $\nu$.
In a mixed model of atoms and molecules, the target distribution function differs between the two because the internal degrees of freedom (DOF) for the molecules must be taken into account.
While the atoms only carry translational DOFs, the molecules carry vibrational and rotational DOFs additionally.
Therefore, the target distribution function of atoms only consists of a translational part
\begin{equation}
	f_{s,\mathrm{Atom}}^t = f_s^{t, \mathrm{tr}}(\mathbf v),
\end{equation}
depending on the particle velocity $\mathbf v$, whereas the distribution function for molecules can be separated into the translational, the rotational and the vibrational part as demonstrated in \citet{MATHIAUD202265,dauvois2021bgk}
\begin{equation}
	f_{s,\mathrm{Molecule}}^t = f_s^{t, \mathrm{tr}}(\mathbf v)f_s^{t,\mathrm{rot}}(E_\mathrm{rot})f_s^{t,\mathrm{vib}}(i_\mathrm{vib}),
\end{equation}
depending also on the rotational energy $E_\mathrm{rot}$ and the vibrational quantum number $i_\mathrm{vib}$.

\subsection{Macroscopic Flow Values}
The macroscopic flow values particle density $n$, flow velocity $u$ and temperature $T_{\mathrm{tr}}$ of each species $s$ are defined as the moments of the particle distribution function:
\begin{alignat}{1}
	&n_s=\int_{\mathbb{R}^3} f_s^{\mathrm{tr}}\,\mathrm d\mathbf v, \quad n_s \mathbf u_s =\int_{\mathbb{R}^3} \mathbf v f_s^{\mathrm{tr}}\,\mathrm d\mathbf v, \\
	&\mathcal{E}_s=\frac{3}{2}k_{\mathrm{B}} T_{\mathrm{tr},s} = \frac{m_s}{2n_s}\int_{\mathbb{R}^3} \mathbf c_s^2 f_s^{\mathrm{tr}}\,\mathrm d\mathbf v,\\
	&E_s=\frac{1}{2}m_sn_s\mathbf u_s^2 + n_s\mathcal{E}_s,
\end{alignat}
with the thermal particle velocity $\mathbf c=\mathbf v -\mathbf u$ from the particle velocity $\mathbf v$, the total energy $E$ and the thermal energy $\mathcal{E}$.
Furthermore, the macroscopic mean values of the flow are given by:
\begin{alignat}{1}
	&\rho=\sum_{s=1}^M m_s n_s,\quad \rho \mathbf u = \sum_{s=1}^M m_s n_s \mathbf u_s,\\
	&n\mathcal{E} + \frac{\rho}{2}\mathbf u^2 = E=\sum_{s=1}^M E_s,
	\quad \mathcal{E}=\frac{3}{2}k_{\mathrm{B}} T_{\mathrm{tr}}.
\end{alignat}
For molecular species, it is also necessary to define the mean rotational energy $\left<E\right>_\mathrm{rot}$ and the mean vibrational energy $\left<E\right>_\mathrm{vib}$, which are directly linked to the rotational temperature $T_\mathrm{rot}$ and the vibrational temperature $T_\mathrm{vib}$:
\begin{alignat}{2}
	&\left<E\right>_{\mathrm{rot},s}&&=\frac{\xi_{\mathrm{rot},s}}{2}k_{\mathrm{B}} T_{\mathrm{rot},s} = \frac{1}{n_s}\int  E_\mathrm{rot} f_s^{\mathrm{rot}}\,\mathrm d E_\mathrm{rot},\\
	&\left<E\right>_{\mathrm{vib},s}&&=\sum_{j=1}^\gamma\frac{\xi_{\mathrm{vib},j,s}}{2}k_{\mathrm{B}} T_{\mathrm{vib},s}\\
	& &&= \frac{1}{n_s}\sum_{j=1}^\gamma\sum_{i_{\mathrm{vib},j}}k_\mathrm{B} \Theta_{\mathrm{vib},j,s} f_s^{\mathrm{vib}}.
\end{alignat}
$\xi$ are the degrees of freedom of rotation and vibration.
In the diatomic case, $\xi_{\mathrm{rot}}=2$ applies, while in the polyatomic case, $\xi_\mathrm{rot}=2$ applies for linear molecules and $\xi_\mathrm{rot}=3$ for non-linear molecules~\cite{relax,herzberg1946infrared}.
For the vibration, the simple harmonic oscillator (SHO) model is assumed with the characteristic vibrational temperature $\Theta_\mathrm{vib}$.
Due to the quantization, the degrees of freedom are not fixed but temperature-dependent.
In the polyatomic case, $\gamma$ is the number of vibration modes as a function of the total number of atoms $a$ in the molecule~\cite{relax}:
\begin{equation}
	\gamma=\begin{cases}
		3a-5,  & \text{linear molecule,}\\
		3a-6, & \text{non-linear molecule.}
	\end{cases}
\end{equation}
In the case of a diatomic molecule, $\gamma=1$ follows.
With the assumption of the SHO, an expression for the mean energy and vibration degrees of freedom as a function of temperature $T_{\mathrm{vib},s}$ can be found:
\begin{alignat}{1}
	&\left<E\right>_{\mathrm{vib},s}=\sum_{j=1}^\gamma\frac{k_\mathrm{B}\Theta_{\mathrm{vib},j,s}}{\mathrm{exp}(\Theta_{\mathrm{vib},j,s}/T_{\mathrm{vib},s})-1},\\
	&\xi_{\mathrm{vib},s} =\sum_{j=1}^\gamma \xi_{\mathrm{vib},j,s} = \sum_{j=1}^\gamma \frac{2\Theta_{\mathrm{vib},j,s}/T_{\mathrm{vib},s}}{\mathrm{exp}(\Theta_{\mathrm{vib},j,s}/T_{\mathrm{vib},s})-1}.
\end{alignat}

\subsection{ESBGK mixture model}
The model proposed here combines the model of \citet{MATHIAUD202265,energy-cons} and the model of \citet{brull, brull2021ellipsoidal} into a mixture model that allows for non-equilibrium effects in the internal degrees of freedom.
Different than the standard BGK model using a Maxwellian target distribution function, the ellipsoidal statistical BGK (ESBGK) model produces the correct Prandtl number by using an anisotropic Gaussian distribution~\cite{esbgk}
\begin{equation}
	f_s^{\mathrm{ES, tr}} = \frac{\rho_s}{\sqrt{\det 2\pi\mathcal A_s}} \exp\left[-\frac{1}{2}\mathbf c^T \mathcal A_s^{-1} \mathbf c\right] \label{eq:esbgk}
\end{equation}
with the mass density of the considered species $\rho_s$ and the average flow velocity $\mathbf u$.
The anisotropic matrix $\mathcal A_s$ reads:
\begin{equation}
	\mathcal A_s = \frac{k_\mathrm{B} T_{\mathrm{tr, rel}}}{m_s} \mathcal I - \frac{1-\alpha Pr}{\alpha Pr}\frac{k_\mathrm{B}}{m_s}\left(\frac{T_{\mathrm{tr, rel}}}{T_{\mathrm{tr}}} \mathcal P-T_{\mathrm{tr, rel}}\mathcal I\right)
\end{equation}
and consists of the identity matrix $\mathcal I$ and the pressure tensor $\mathcal P$
\begin{equation}
	\mathcal P =\frac{1}{\rho}\sum_{s=1}^M m_s\int (\mathbf v-\mathbf u) (\mathbf v-\mathbf u)^T f_s\,\mathrm d\mathbf v,
\end{equation}
which are both symmetric~\cite{brull}.
The parameter $\alpha$ is a variable of the model depending on mass fraction, density fraction and internal degrees of freedom~\cite{brull2021ellipsoidal}:
\begin{align}
	&\alpha = m \frac{\sum_{s=1}^M \frac{n_s}{m_s} \left(5 + \xi_{\mathrm{int},s}\right)}{\sum_{s=1}^M n_s \left(5 + \xi_{\mathrm{int},s}\right)},\\
	&m=\sum_{s=1}^M \frac{n_s}{n}m_s,
	\quad n=\sum_{s=1}^M n_s. \label{eq:m}
\end{align}
The internal degrees of freedom are calculated as the sum of the vibrational and rotational degrees of freedom, respectively:
\begin{equation}
	\xi_{\mathrm{int},s} = \xi_{\mathrm{vib},s}+\xi_{\mathrm{rot},s}.
	\label{eq:intDOF}
\end{equation}
The model used and presented in this paper uses one relaxation term per species $s$.
It fulfills the indifferentiability principle so that the model reduces to a single species model for identical species.
Additionally, the Maxwellian distribution is produced in the equilibrium state.
The relaxation frequency $\nu$ of the model is defined by 
\begin{equation}
	\nu = \frac{nk_{\mathrm{B}} T_{\mathrm{tr}}}{\mu}\alpha Pr \label{eq:nu}
\end{equation}
with the viscosity of the mixture $\mu$ and the translational temperature $T_{\mathrm{tr}}$.
$Pr$ is the targeted Prandtl number of the gas mixture, which is calculated using Wilke's mixing rules~\cite{wilke} or collision integrals~\cite{collint} (see Section~\ref{sec:mixture}).

For molecular species, the rotational and vibrational components of the distribution function must also be defined, which are taken directly from \citet{MATHIAUD202265}:
\begin{alignat}{1}
	f_s^{\mathrm{ES,rot}} = &\frac{E_\mathrm{rot}^{(\xi_{\mathrm{rot},s}-2)/2}}{\Gamma ( \xi_{\mathrm{rot},s}/2)\left(k_\mathrm{B} T_{\mathrm{rot, rel,s}}\right)^{\xi_{\mathrm{rot},s}/2}}\nonumber\\
	&\cdot\exp\left[-\frac{E_\mathrm{rot}}{k_\mathrm{B} T_{\mathrm{rot, rel,s}}}\right],\label{eq:distrot}\\
	f_s^{\mathrm{ES,vib}} = &\prod_{j=1}^\gamma \left(1-\exp\left[-\frac{\Theta_{\mathrm{vib},j,s}}{T_{\mathrm{vib, rel},s}}\right]\right) \nonumber\\
	&\cdot\exp\left[-i_j\frac{\Theta_{\mathrm{vib},j,s}}{T_{\mathrm{vib, rel},s}}\right]\label{eq:distvib}.
\end{alignat}
The temperatures $T_{\mathrm{tr, rel}}$, $T_{\mathrm{rot,rel},s}$ and $T_{\mathrm{vib,rel},s}$ introduced here are chosen so that the relaxation corresponds to the Landau-Teller relaxation:
\begin{alignat}{1}
	&\frac{\mathrm{d}\left<E\right>_{r,s}(T_{r,s})}{\mathrm{d}t} = \frac{1}{Z_{r,s}\tau_{s,\mathrm{C}}}\left(\left<E\right>_{r,s}(T_{\mathrm{tr}}) - \left<E\right>_{r,s}(T_{r,s})\right),\nonumber\\
	& r = \mathrm{rot},\, \mathrm{vib}.
\end{alignat}
Here $Z_{r,s}$ is the collision number of the internal degrees of freedom $r$ per species $s$ and $\tau_{s,\mathrm{C}}=1/\nu_{s,\mathrm{C}}$ is the collision time of the gas per species.
The latter is different from $\tau=1/\nu$, the relaxation time of the BGK equation~\eqref{eq:BGK}, whereas a detailed discussion can be found in \citet{MATHIAUD202265}.
The collision frequency for the variable hard sphere model (VHS) of each species $s$ in a mixture of $M$ species is calculated by~\cite{chapman-cowling}:
\begin{equation}
	\nu_{s,\mathrm{C}}= \sum_{k=1}^M 2 d_{s,k}^2 
	n_k \sqrt{\frac{2 \pi k_{\mathrm{B}} T_{s,k}} {\left(m_s+m_k\right)} {m_s m_k}} \left(\frac{T_{s,k}}{T_{\mathrm{tr}}}\right)^{\omega_{s,k}}
\end{equation}
with	
\begin{align}
	&d_{s,k} = 0.5 \left(d_{\mathrm{VHS},s}+d_{\mathrm{VHS},k}\right), \\
	&T_{s,k} = 0.5 \left(T_{\mathrm{VHS},s}+T_{\mathrm{VHS},k}\right), \\
	&\omega_{s,k} = 0.5 \left(\omega_{\mathrm{VHS},s}+\omega_{\mathrm{VHS},k}\right).
\end{align}
If direct VHS data are available for each collision combination, these can also be used accordingly here~\cite{swaminathan2016recommended,10.1063/5.0118040,hong2023optimized}.
With these definitions, $T_{\mathrm{tr, rel}}$, $T_{\mathrm{rot,rel},s}$ and $T_{\mathrm{vib,rel},s}$ can now be defined with $r = \mathrm{rot},\, \mathrm{vib}$:
\begin{alignat}{1}
	&\left<E\right>_{r,s}(T_{r, \mathrm{rel},s})= \left<E\right>_{r,s}(T_{r,s}) \nonumber\\
	&+\frac{\tau}{Z_{r,s}\tau_{s,\mathrm{C}}}\left(\left<E\right>_{r,s}(T_{\mathrm{tr}}) - \left<E\right>_{r,s}(T_{r,s}) \right),\label{eq:newtempinner}\\
	&\left<E\right>_\mathrm{tr}(T_{\mathrm{tr, rel}}) = \left<E\right>_\mathrm{tr}(T_{\mathrm{tr}}) \nonumber\\ 
	&- \sum_{r}\sum_{s}^{M_\mathrm{molec}} \frac{\tau}{Z_{r,s}\tau_{s,\mathrm{C}}}\left(
	\left<E\right>_{r,s}(T_{\mathrm{tr}}) - \left<E\right>_{r,s}(T_{r,s})  \right).
\end{alignat}

\subsection{Gas mixture properties} \label{sec:mixture}
The ESBGK model requires the calculation of the correct Prandtl number
\begin{equation}
	Pr = c_{\mathrm{p}} \frac{\mu}{K}
\end{equation}
of the gas mixture to assess the relaxation frequency from Equation~\eqref{eq:nu}.
Here, $\mu$ is the viscosity, $K$ is the thermal conductivity and $c_{\mathrm{p}}$ is the specific heat of the considered mixture.
The latter is calculated as:
\begin{equation}
	c_{\mathrm{p}} = \sum_{s=1}^M \frac{5+\xi_{\mathrm{int},s}}{2}k_{\mathrm{B}}\frac{n_s}{\sum_{k=1}^M n_k m_k}.           
\end{equation}
For the calculation of the mixture viscosity and thermal conductivity, any source can be used in the presented model.
Basically, the same problems also arise with multi-species CFD codes and therefore the same solutions can also be applied.
Typically, the transport properties are calculated on the basis of Wilke’s mixture rules~\cite{wilke} or the first approximation of the transport properties using collision integrals~\cite{collint}, whereby various libraries are also available that can be used directly~\cite{Scoggins2020,kee1986fortran,ern2004eglib}.
However, in order to achieve the same transport coefficients as far as possible when coupling with DSMC, the procedure for the Variable Hard Sphere (VHS) potential model is presented in this paper.

\subsubsection{Wilke's mixing rules}
Wilke's mixing rules~\cite{wilke} are used for the calculation of the transport coefficients of the mixture.
The well-known exponential ansatz is used for the viscosity of each species $\mu_s$, which can be calculated with:
\begin{equation}
	\mu_s = \mu_{\mathrm{ref},s} \left( \frac{T_{\mathrm{tr}}}{T_{\mathrm{ref},s}} \right) ^{\omega_{\mathrm{VHS},s}}.
\end{equation}
Here, $\mu_{\mathrm{ref},s}$ is the reference dynamic viscosity at a reference temperature $T_{\mathrm{ref},s}$ for each species $s$, and $\omega_{\mathrm{VHS},s}$ is a species-specific parameter of the Variable Hard Sphere (VHS) model~\cite{bird,10.1063/5.0118040,swaminathan2016recommended,hong2023optimized}.
Using the VHS reference diameter $d_{\mathrm{ref},s}$, the reference dynamic viscosity for a VHS gas is defined as~\cite{burt-boyd}:
\begin{equation}
	\mu_{\mathrm{ref},s}=\frac{30 \sqrt{m k_{\mathrm{B}} T_{\mathrm{ref},s}}}{\sqrt{\pi} 4 (5-2\omega_{\mathrm{VHS},s}) (7-2\omega_{\mathrm{VHS},s}) d_{\mathrm{ref},s}^2}.
\end{equation}
The mixture viscosity of $M$ species is then calculated by~\cite{wilke}:
\begin{equation}
	\mu = \sum_{s=1}^M n_s \frac{\mu_s}{\Phi_s},
	\quad \Phi_s = \sum_{k=1}^M n_k \frac{\left(1+\sqrt{\frac{\mu_s}{\mu_k}}\left(\frac{m_k}{m_s}\right)^{1/4}\right)^2}{\sqrt{8\left(1+\frac{m_s}{m_k}\right)}}.
\end{equation}
The thermal conductivity of each species $K_s$ can be calculated using the Eucken's relation~\cite{eucken} and the correction by Hirschfelder\cite{hirschfelder1957heat} with the viscosity:
\begin{equation}
	K_s = (f_{\mathrm{tr}} c_{\mathrm{v,tr},s}+ f_{\mathrm{int}} c_{\mathrm{v,int},s}) \mu_s. \label{eq:K_s}
\end{equation}
Here, $c_{\mathrm{v},s}=\xi_s k_{\mathrm{B}}/2m_s$ is the specific heat capacity at constant volume for species $s$, with the corresponding degrees of freedom for translation $\xi_{\mathrm{tr}}=3$ for each species and the internal degrees of freedom $\xi_{\mathrm{int},s}$ from Equation~\eqref{eq:intDOF}.
In general, $c_{\mathrm{v}}=c_{\mathrm{p}}-k_{\mathrm{B}}/m$ applies.
The prefactors used in Equation~\eqref{eq:K_s} are defined as $f_{\mathrm{tr}}=5/2$ and $f_{\mathrm{int}}=1.328$.
With that, the thermal conductivity of the mixture $K$ is calculated:
\begin{equation}
	K=\sum_{s=1}^M n_s \frac{K_s}{\Phi_s}.
\end{equation}

\subsubsection{First approximation of transport properties}
Another calculation possibility is the first approximation to the viscosity of species $s$ depending on the collision integral $\Omega_s^{(2)}(2)$, which is given by~\cite{collint}
\begin{equation}
	\mu_s=\frac{5k_{\mathrm{B}}T_{\mathrm{tr},s}}{8\Omega_s^{(2)}(2)}.
\end{equation}
The mixture viscosity is determined by 
\begin{equation}
	\mu=\sum_{s=1}^M b_s,
\end{equation}
where $b_s$ is the contribution of each species to the total mixture viscosity and is determined by solving the system
\begin{align}
	\chi_s &= b_s\left(\frac{\chi_s}{\mu_s} + \sum_{k\neq s}\frac{3\chi_s}{(\rho_k'+\rho_s')D_{sk}}\left(\frac{2}{3}+\frac{m_k}{m_s}A_{sk}\right)\right)\nonumber\\
	&-\chi_s\sum_{k\neq s}\frac{3 b_k}{(\rho_k'+\rho_s')D_{sk}}\left(\frac{2}{3}-A_{sk}\right)
\end{align}
with the mole fraction $\chi$, the density $\rho_k'$ of species $k$ when pure at pressure and temperature of the actual gas mixture, the parameter $A_{sk}$, defined by
\begin{equation}
	A_{sk}=\frac{\Omega_{sk}^{(2)}(2)}{5\Omega_{sk}^{(1)}(1)},
\end{equation}
and the binary diffusion coefficient with the reduced mass $m^*_{sk}$:
\begin{equation}
	D_{sk}=\frac{3k_{\mathrm{B}} T_{\mathrm{tr}}}{16nm^*_{sk}\Omega_{sk}^{(1)}(1)}.
\end{equation}
The mixture thermal conductivity $K$ is calculated by
\begin{equation}
	K=\sum_{s=1}^M a_s + \sum_{s=1}^{M_\mathrm{molec}} K_{\mathrm{rot},s} + \sum_{s=1}^{M_\mathrm{molec}} K_{\mathrm{vib},s},
\end{equation}
with $a_s$ being the translational species contribution to the total mixture thermal conductivity and $K_{\mathrm{rot},s}$ as well as $K_{\mathrm{vib},s}$ being the rotational and vibrational contributions of the species $s$~\cite{chapman-cowling}.
The factors $a_s$ are determined by solving the system
\begin{align}
	\chi_s &= a_s\left[\frac{\chi_s}{K_s} + \sum_{k\neq s}\frac{\chi_k}{5k_{\mathrm{B}} n D_{sk}} \cdot\left(6\left(\frac{m_s}{m_k + m_s}\right)^2\right.\right.\nonumber\\
	&\left.\left. + (5-4B_{sk})\left(\frac{m_k}{m_k + m_s}\right)^2 + 8\frac{m_k m_s}{(m_s+m_k)^2}A_{sk}\right)\right]\nonumber\\
	&-\chi_s\sum_{k\neq s}a_k\frac{m_k m_s}{(m_s+m_k)^2}(5k_{\mathrm{B}}n D_{sk})^{-1}\nonumber \\
	&\cdot\left(11-4B_{sk}-8A_{sk}\right).
\end{align}
Here, $K_s$ is the first approximation of the thermal conductivity of species $s$:
\begin{equation}
	K_s=\frac{25 c_{\mathrm{v,tr},s} k_{\mathrm{B}} T_{\mathrm{tr},s}}{16 \Omega_s^{(2)}(2)},\quad c_{\mathrm{v,tr},s}=\frac{3k_{\mathrm{B}}}{2m_s},
\end{equation}
and the parameter $B_{sk}$ is defined by
\begin{equation}
	B_{sk}=\frac{5\Omega_{sk}^{(1)}(2)-\Omega_{sk}^{(1)}(3)}{5\Omega_{sk}^{(1)}(1)}.
\end{equation}
The collision integrals for the VHS model are given by~\cite{Stephani2012}
\begin{align}
	\Omega_{sk}^{\mathrm{VHS},(1)}(1)&=\frac{\pi}{2}d_{\mathrm{ref}}^2\sqrt{\frac{k_{\mathrm{B}}T_{\mathrm{tr}}}{2\pi m^*_{sk}}}\left(\frac{T_{\mathrm{ref}}}{T_{\mathrm{tr}}}\right)^{\omega-1/2}\frac{\Gamma(7/2-\omega)}{\Gamma(5/2-\omega)}\nonumber\\
	\Omega_{sk}^{\mathrm{VHS},(2)}(2)&=\frac{\pi}{3}d_{\mathrm{ref}}^2\sqrt{\frac{k_{\mathrm{B}}T_{\mathrm{tr}}}{2\pi m^*_{sk}}}\left(\frac{T_{\mathrm{ref}}}{T_{\mathrm{tr}}}\right)^{\omega-1/2}\frac{\Gamma(9/2-\omega)}{\Gamma(5/2-\omega)}\nonumber\\
	B^{\mathrm{VHS}}_{sk}&=\frac{5\Gamma(9/2-\omega)-\Gamma(11/2-\omega)}{5\Gamma(7/2-\omega)}
\end{align}
with the VHS parameters $d_{\mathrm{ref}}$, $T_{\mathrm{ref}}$ and $\omega$.
Generally, for $s=k$, the translational temperature of the species $T_{\mathrm{tr},s}$ is used, while for $s \neq k$, the translational temperature in the cell $T_{\mathrm{tr}}$ is used as an approximation in order to reduce noise.

The rotational and vibrational contributions $K_{r,s}$ with $r=\mathrm{rot,vib}$ to the thermal conductivity of a species $s$ are calculated as~\cite{chapman-cowling}:
\begin{equation}
	K_{r,s} = \frac{n_s m_s c_{\mathrm{v},r,s}}{\sum_{k=1}^{M} \chi_k D_{sk}^{-1}}, \quad c_{\mathrm{v},r,s} = \frac{\xi_{r,s} k_{\mathrm{B}}}{2m_s}.
\end{equation}

\section{Implementation}
The proposed ESBGK method gas mixtures with internal degrees of freedom is implemented in the open-source particle code PICLas~\cite{piclas}.
A stochastic particle approach is chosen for the implementation~\cite{piclas-bgk,energy-cons,bgk-multispecies,bgk-poly}, although the model can of course also be used in discrete velocity (DVM) codes.
Especially to save simulation time in DVM multi-species simulations, the model should be extended to a reduced model in the future~\cite{chu1965kinetic, dauvois2021bgk, MATHIAUD202265}.

\subsection{Relaxation and sampling}
Regarding movement and boundary conditions, particles are treated in the same manner as in the DSMC method.
But instead of performing binary collisions between the particles, all particles in a cell relax with a probability $P$ in each time step $\Delta t$:
\begin{equation}
	P = 1 - \exp\left[-\nu \Delta t\right], \label{eq:prob}
\end{equation}
which corresponds to Equation~\eqref{eq:BGK} in a stochastic interpretation as described in \citet{piclas-bgk}.
The first order method from \citet{piclas-bgk} is used here.
In the future, however, this model will be extended to the second-order particle method from \citet{PhysRevE.106.025303}.
The relaxation frequency $\nu$, which is the same as in Equation~\eqref{eq:BGK}, is evaluated in each time step for each cell according to Equation~\eqref{eq:nu} using the transport coefficients of the mixture calculated in the code.

The particles that are chosen to relax are then sampled from the target distribution (see Equation~\eqref{eq:esbgk}).
Here, an approximation of the transformation matrix $\mathcal{S}$~\cite{gallis-torczynski} with $\mathcal{A}=\mathcal{S}\mathcal{S}$ is used to transform a random vector chosen from a Maxwellian to a velocity vector chosen from the ESBGK distribution:
\begin{equation}
	v_p^* = u + \mathcal{S} \sqrt{R_p\frac{k_\mathrm{B} T_\mathrm{tr,rel}}{m_s}}.
\end{equation}
For a detailed description of the sampling process, the reader is referred to Ref.~\cite{piclas-bgk}, in which also alternative methods for sampling are described.

At this point, a special solution for the stochastic particle approach must be chosen.
Even though the presented model is both momentum and energy conserving, additional effort must be made for both in a stochastic approach.
The basic problem is that the new states of the particles (velocities, internal energies) are chosen randomly from the distribution functions.
Thus, in this random step, conservation of energy and momentum is not automatically given and one would need a large number of particles for stable simulations.
To construct energy conservation for the stochastic particle method, the energy difference between the old and new sampled energy must be distributed over the particles in such a way that energy conservation is given.
The idea is to distribute this energy difference evenly according to the respective degrees of freedom to the translational energy of all particles in the cell and the internal degrees of freedom of the particles relaxing in the degrees of freedom within the cell, as this has led to the best results in previous work~\cite{energy-cons}. 
An artificial equilibrium temperature $T_{\mathrm{equi}}$  is introduced for this purpose.
Starting from this point, molecules are chosen to relax in the rotation and vibration towards the equilibrium temperature $T_{\mathrm{equi}}$ with the probability
\begin{equation}
	P_{r,s} = P \beta_{r,s} \frac{\tau}{Z_{r,s}\tau_{s,\mathrm{C}}}, \quad r = \mathrm{rot, vib}, \label{eq:P-internal}
\end{equation}
so that Equations~\eqref{eq:BGK}, \eqref{eq:distrot} and \eqref{eq:distvib} are fulfilled.
To save calculation time, instead of using the probability $P$ to relax the particles to $T_{\mathrm{vib, rel},s}$ and $T_{\mathrm{rot, rel},s}$ corresponding to Equation~\eqref{eq:distrot} and \eqref{eq:distvib}, only the fraction $\frac{\tau}{Z_{r, s}\tau_{s,\mathrm{C}}}$ relaxes directly to $T_{\mathrm{tr}}$ according to Equation~\eqref{eq:newtempinner}, which leads to the same result but with significantly fewer relaxing particles.
Additionally, a correction parameter $\beta_{r,s}$ for the rotation and vibration, respectively, is defined per species in order to follow the idea of equal distribution of energies for the conservation of energy as described:
\begin{align}
	&\frac{\tau}{Z_{r,s}\tau_{s,\mathrm{C}}} \left(\left<E\right>_{r,s}(T_{\mathrm{tr}})-\left<E\right>_{r,s}(T_{r,s})\right) \nonumber\\
	&= \beta_{r,s} \frac{\tau}{Z_{r,s}\tau_{s,\mathrm{C}}} \left(\left<E\right>_{r,s}(T_{\mathrm{equi}})-\left<E\right>_{r,s}(T_{r,s})\right) \nonumber\\
	&\Rightarrow \beta_{r,s} = \frac{T_{r,s} - T_{\mathrm{tr}}}{T_{r,s} - T_{\mathrm{equi}}}.
\end{align}
Since $T_\mathrm{equi}$ directly depends on $\beta_{\mathrm{rot},s}$ and $\beta_{\mathrm{vib},s}$ of each species, the solution of this equation system is obtained numerically.
For this purpose, the following system must be solved:
\begin{align}
	T_\mathrm{equi}^{n+1} = &\left[3 (N-1) T_\mathrm{tr} + \sum_{s=1}^{M_\mathrm{molec}} \xi_{\mathrm{rot},s} N_s P_{\mathrm{rot},s} (\beta_{\mathrm{rot},s}^n) T_{\mathrm{rot},s} \right. \nonumber\\
	&\left.+ \sum_{s=1}^{M_\mathrm{molec}} \xi_{\mathrm{vib},s} N_s P_{\mathrm{vib},s} (\beta_{\mathrm{vib},s}^n) T_{\mathrm{vib},s}\right] \nonumber\\
	&/ \left[3 (N-1) + \sum_{s=1}^{M_\mathrm{molec}} \xi_{\mathrm{rot},s} N_s P_{\mathrm{rot},s} (\beta_{\mathrm{rot},s}^n) \right. \nonumber\\
	&\left.+ \sum_{s=1}^{M_\mathrm{molec}} \xi_{\mathrm{vib},s}(T_{\mathrm{equi}}^n) N_s P_{\mathrm{vib},s} (\beta_{\mathrm{vib},s}^n)\right],\\
	\beta_{r,s}^{n+1} = &\frac{T_{r,s} - T_{\mathrm{tr}}}{T_{r,s} - T_{\mathrm{equi}}^{n+1}},
\end{align}
with the starting values
\begin{align}
	\beta_{r,s}^0 &= 1,\\
	\xi_{\mathrm{vib},s}(T_{\mathrm{equi}}^0) &= \xi_{\mathrm{vib},s}(T_{\mathrm{vib},s}),
\end{align}
and iteration step $n$, until the accuracy $\epsilon$ is reached with the condition $T_\mathrm{equi}^{n+1} -T_\mathrm{equi}^n < \epsilon$.
$N$ is hereby defined as the total number of particles in the gas mixture, while $N_s$ is the particle number per species.

The new rotational energy of a molecule $p$, which is chosen for a rotational relaxation according to the probability in Equation~\eqref{eq:P-internal}, is calculated by:
\begin{equation}
	E_{\mathrm{rot},p}^* = - \frac{\xi_{\mathrm{rot},s}}{2} k_\mathrm{B} T_\mathrm{equi} \mathrm{ln}(R_p),
\end{equation}
using a random number $R_p$.
In the same manner, the new vibrational energy of a vibrational relaxing particle is calculated:
\begin{equation}
	E_{\mathrm{vib},p}^* = - \sum_{j=1}^{\gamma} \frac{\xi_{\mathrm{vib},j,s}(T_\mathrm{equi})}{2} k_\mathrm{B} T_\mathrm{equi} \mathrm{ln}(R_{p,j}).
\end{equation}
The post-relaxation energies are hereby denoted with $^*$.
To fulfill the energy conservation, these energies need to be scaled additionally.
Note: At this point the vibration is still continuous but already with the correct quantized degrees of freedom.
The transition to quantized states happens with the conservation of energy.

\subsection{Energy and momentum conservation}
The energy and momentum conservation is based on Ref.~\cite{energy-cons} and extended to polyatomic molecules and mixtures of different gas species.
In the implemented scheme, only relaxations in translation-vibration ($T-V$) and translation-rotation ($T-R$) are allowed directly.
In the following, the process is described in greater detail.
Here, the parameters after energy conservation are denoted with $'$.

For the energy conservation process, the vibrational energy is scaled first.
The energy $E_{T-V}$ is the sum of the total translational energy of the gas mixture and the vibrational energy of all relaxing particles in vibration, $N_\mathrm{vib}$, before the relaxation process:
\begin{equation}
	E_{T-V} = \sum_{p=1}^{N} \frac{m}{2} c_p^2 + \sum_{p=1}^{N_\mathrm{vib}} \left( E_{\mathrm{vib},p} - \sum_{j=1}^{\gamma} 0.5 k_\mathrm{B} \Theta_{\mathrm{vib},j,s} \right).
\end{equation}
It should be distributed equally over all translational and vibrational DOFs after the relaxation process to fulfill the energy conservation.
This is reached with a scaling factor $\alpha_{\mathrm{vib},s}$, that is calculated per species as:
\begin{align}
	\alpha_{\mathrm{vib},s} = &\frac{E_{T-V}}{\sum_{p=1}^{N_\mathrm{vib}} E_{\mathrm{vib},p,s}^*} \nonumber\\
	&\cdot\left( \frac{\xi_{\mathrm{vib},s}(T_\mathrm{equi}) N_{\mathrm{vib},s}}{3 (N-1) + \sum_{s=1}^{M_\mathrm{molec}} \xi_{\mathrm{vib},s}(T_\mathrm{equi}) N_{\mathrm{vib},s}} \right).
\end{align}
With this, the new vibrational energy of each particle $p$ after the relaxation and energy conservation processes is
\begin{equation}
	E_{\mathrm{vib},p}' = \alpha_{\mathrm{vib},s} E_{\mathrm{vib},p}^* + \sum_{j=1}^{\gamma} 0.5 k_\mathrm{B} \Theta_{\mathrm{vib},j,s}
\end{equation}
for the continuous handling of the vibrational energies and including the zero-point energy corresponding to the second sum.
For quantized vibrational energy states, additional steps are to be performed.
The energy per vibrational mode $\alpha_{\mathrm{vib},s} E_{\mathrm{vib},p,j}^*$ is reformulated to a quantum number $i_{p,j}$ with a random number $R_{p,j}\in [0,1)$:
\begin{equation}
	i_{p,j} = \text{INT} \left( \frac{\alpha_{\mathrm{vib},s} E_{\mathrm{vib},p,j}^*}{k_\mathrm{B} \Theta_{\mathrm{vib},j,s}} +R_{p,j} \right).
\end{equation}
If the condition
\begin{equation}
	E_{T-V} > i_{p,j} k_\mathrm{B} \Theta_{\mathrm{vib},j,s} \label{eq:energy-cons-vib-quant-cond}
\end{equation}
is fulfilled,
\begin{equation}
	E_{\mathrm{vib},p,j}' = (i_{p,j}+0.5) k_\mathrm{B} \Theta_{\mathrm{vib},j,s} \label{eq:energy-vib-new}
\end{equation}
is the new energy of this vibrational mode, including the zero-point energy.
Otherwise, a new quantum number of this mode is calculated with a new random number $R_{p,j}$ as:
\begin{equation}
	i_{p,j} = \text{INT} \left( -\mathrm{ln}(R_{p,j}) \frac{T_\mathrm{equi}}{\Theta_{\mathrm{vib},j,s}} \right)
\end{equation}
until Equation~\eqref{eq:energy-cons-vib-quant-cond} is fulfilled.
$E_{\mathrm{vib},p,j}'$ is then calculated with this new quantum number using Equation~\eqref{eq:energy-vib-new} subsequently.
The energy $E_{T-V}$ is subsequently updated with
\begin{equation}
	E_{T-V} = E_{T-V} - i_{p,j} k_\mathrm{B} \Theta_{\mathrm{vib},j,s},
\end{equation}
before the algorithm is repeated for each vibrational mode $j$ of each vibrational relaxing particle $p$.
In the end, the new energy of the each particle $p$ is:
\begin{equation}
	E_{\mathrm{vib},p}' = \sum_{j=1}^{\gamma} E_{\mathrm{vib},p,j}'.
\end{equation}

Afterwards, the remaining energy for translation and rotation is the sum of the remaining $E_{T-V}$ and the rotational energy of all rotational relaxing particles before the relaxation process:
\begin{equation}
	E_{T-R} = E_{T-V} + \sum_{p=1}^{N_\mathrm{rot}} E_{\mathrm{rot},p}.
\end{equation}
Here, $N_\mathrm{rot}$ is the number of the rotational relaxing particles.
$E_{T-R}$ should be distributed equally over all translational and rotational DOFs similar to $E_{T-V}$.
For this, a scaling factor $\alpha_{\mathrm{rot},s}$ is calculated per species as:
\begin{align}
	\alpha_{\mathrm{rot},s} = &\frac{E_{T-R}}{\sum_{p=1}^{N_\mathrm{rot}} E_{\mathrm{rot},p,s}^*} \nonumber\\
	&\cdot\left( \frac{\xi_{\mathrm{rot},s} N_{\mathrm{rot},s}}{3 (N-1) + \sum_{s=1}^{M_\mathrm{molec}} \xi_{\mathrm{rot},s} N_{\mathrm{rot},s}} \right).
\end{align}
Using this, the new rotational energy of each particle $p$ after the relaxation and energy conservation processes is calculated with:
\begin{equation}
	E_{\mathrm{rot},p}' = \alpha_{\mathrm{rot},s} E_{\mathrm{rot},p}^*.
\end{equation}

Different than for the internal DOFs, and as described earlier, all particles in the gas mixture are scaled in the translation, whether they relax or not.
The scaling factor $\alpha_\mathrm{tr}$ is derived to be
\begin{align}
	\alpha_\mathrm{tr} = &\left[\frac{E_{T-R}}{\sum_{p=1}^{N} \frac{m}{2} (v_p^*-u^*)^2} \right. \nonumber\\
	&\left.\cdot \left(  \frac{3 (N-1)}{3 (N+1) + \sum_{s=1}^{M_\mathrm{molec}} \xi_{\mathrm{rot},s} N_{\mathrm{rot},s}} \right)\right]^{1/2}.
\end{align}
Finally, the new particle velocities are calculated using
\begin{equation}
	v_p' = u + \alpha_\mathrm{tr} \left( v_p^* - u^* \right),
\end{equation}
with the average flow velocities
\begin{equation}
	u = \sum_{p=1}^{N} v_p/N, \quad u^* = \sum_{p=1}^{N} v_p^*/N.
\end{equation}
If no relaxation occurs for a particle $p$, $v_p^*=v_p$ applies. Using this approach, energy conservation is guaranteed and
momentum conservation is ensured due to
\begin{equation}
	\sum_{p=1}^{N} \left( v_p^* - u^* \right) = 0
\end{equation}
as described in~\citet{energy-cons}.

\section{Simulation Results}
The previously described multi-species ESBGK implementation including molecules with internal degrees of freedom is verified with simulation test cases of a supersonic Couette flow as well as a hypersonic flow around a 70° blunted cone.
The different approaches for the calculation of the transport coefficients, Wilke’s mixing rules (denoted by Wilke) and the direct use of collision integrals (denoted by CollInt), are compared using the VHS model.
The VHS species-specific parameters are listed in Table~\ref{t:VHS}.
The parameters for a collision pair of unlike species are determined as an average (referred to as collision-averaged).
\begin{table}
	\centering
	\renewcommand{\arraystretch}{1.3}
	\begin{tabular}{l|l|l|l}
		Gas species & $d_\mathrm{VHS}$ / $\si{\meter}$ & $T_\mathrm{VHS}$ / $\si{\kelvin}$ & $\omega_\mathrm{VHS}$ / $-$ \\
		\hline
		$\mathrm{He}$ & $\num{2.33e-10}$ & $273$ & $0.77$ \\
		$\mathrm{Ar}$ & $\num{4.05e-10}$ & $273$ & $0.77$ \\
		$\mathrm{N}$ & $\num{3.00e-10}$ & $273$ & $0.74$ \\
		$\mathrm{O}$ & $\num{3.00e-10}$ & $273$ & $0.74$ \\
		$\mathrm{N}_2$ & $\num{4.17e-10}$ & $273$ & $0.74$ \\
		$\mathrm{O}_2$ & $\num{3.98e-10}$ & $273$ & $0.74$ \\
		$\mathrm{NO}$ & $\num{4.00e-10}$ & $273$ & $0.74$ \\
		$\mathrm{CO}_2$ & $\num{5.10e-10}$ & $273$ & $0.74$ \\
	\end{tabular}
	\caption{VHS species-specific parameters~\cite{bird}.}
	\label{t:VHS}
\end{table}

\subsection{Supersonic Couette flow}
The first test case is a three-dimensional simulation of a supersonic Couette flow.
A mesh with 100 cells in $y$ direction and a single cell in $x$ and $z$ directions is used.
For the second test case of $\mathrm{N}_2$-$\mathrm{N}$ with $Kn_\mathrm{VHS} \approx 0.121$ and $Kn_\mathrm{VHS} \approx 1.21$ as well as for test cases four and five, the mesh is refined and thus contains 400 cells in $y$ direction.
The time steps and particle weighting factors are chosen so that the mean free path and the collision frequency are resolved.
The boundaries in $y$ direction have a velocity of $v_\mathrm{wall,1}=\SI{350}{\meter\per\second}$ and $v_\mathrm{wall,2}=\SI{-350}{\meter\per\second}$, respectively, assuming diffuse reflection and complete thermal accommodation at a constant wall temperature of $T_\mathrm{wall}=\SI{273}{K}$.
In $x$ and $z$ directions, all boundaries are periodic, leading to particles reappearing on a side after leaving the opposite side.
Different gas mixtures are simulated, all initialized at $v_0=\SI{0}{\meter\per\second}$ and $T_0=\SI{273}{K}$.
An overview of all Couette flow test cases is given in Table~\ref{t:couette}.
\begin{table}
	\centering
	\renewcommand{\arraystretch}{1.3}
	\begin{tabular}{l|l|p{2cm}|p{1.5cm}}
		& Gas species & Composition & $n_0$ / $\si{\per\cubic\meter}$ \\
		\hline
		Case 1 & $\mathrm{N}_2$-$\mathrm{He}$ & $50\:\%$-$50\:\%$,\newline $20\:\%$-$80\:\%$,\newline $80\:\%$-$20\:\%$ & $\num{1.3e20}$ \\
		Case 2 & $\mathrm{N}_2$-$\mathrm{N}$ & $50\:\%$-$50\:\%$ & $\num{1.3e20}$,\newline $\num{1.3e19}$,\newline $\num{1.3e18}$ \\
		Case 3 & $\mathrm{CO}_2$-$\mathrm{N}_2$ & $50\:\%$-$50\:\%$ & $\num{1.3e20}$ \\
		Case 4 & $\mathrm{N}_2$-$\mathrm{O}_2$-$\mathrm{Ar}$ & $78\:\%$-$21\:\%$-$1\:\%$ & $\num{1.0e20}$ \\
		Case 5 & $\mathrm{N}$-$\mathrm{O}$-$\mathrm{N}_2$-$\mathrm{O}_2$-$\mathrm{NO}$ & $20\:\%$ each & $\num{1.25e20}$ \\
	\end{tabular}
	\caption{Overview of Couette flow test cases.}
	\label{t:couette}
\end{table}

\begin{figure}
	\centering
	\includegraphics{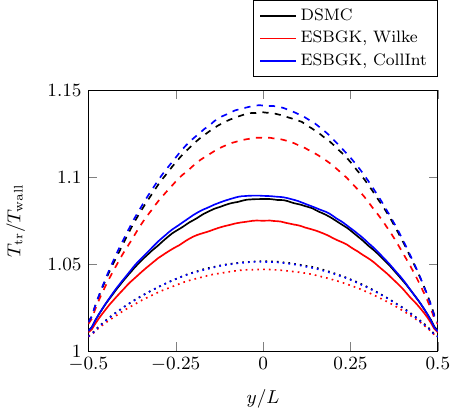}
	\caption{Comparison of the stationary solution for the translational temperature in a supersonic Couette flow for a $\mathrm{N}_2$-$\mathrm{He}$ mixture. The $50\:\%$-$50\:\%$ mixture results are indicated with solid lines, the $80\:\%$-$20\:\%$ mixture with dashed lines and the $20\:\%$-$80\:\%$ mixture with dotted lines.}
	\label{fig:couette_N2-He}
\end{figure}

\begin{figure}
	\centering
	\includegraphics{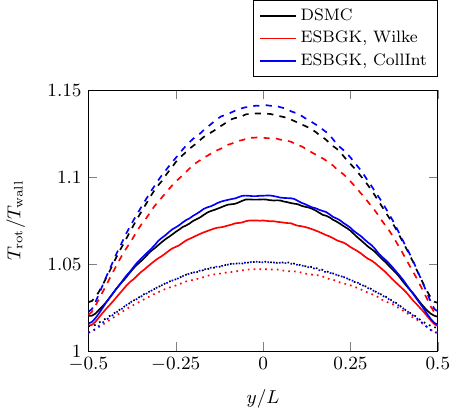}
	\caption{Comparison of the stationary solution for the rotational temperature in a supersonic Couette flow for a $\mathrm{N}_2$-$\mathrm{He}$ mixture. The $50\:\%$-$50\:\%$ mixture results are indicated with solid lines, the $80\:\%$-$20\:\%$ mixture with dashed lines and the $20\:\%$-$80\:\%$ mixture with dotted lines.}
	\label{fig:couette_N2-He-rot}
\end{figure}

First, a $\mathrm{N}_2$-$\mathrm{He}$ mixture is simulated, which represents a challenging case due to the relatively large mass differences.
Mixing ratios of $50\:\%$-$50\:\%$, $20\:\%$-$80\:\%$ and $80\:\%$-$20\:\%$ with a initial particle density of $n_0=\SI{1.3e20}{\per\cubic\meter}$ are tested to estimate the influence of composition on the simulation results.
The correct Prandtl number approximation is crucial for an accurate temperature representation in the Couette case.
However, since it is not a linear function of the mixing ratio~\cite{giacobbe1994estimation}, these three ratios were chosen to capture the resulting Prandtl numbers as a function of the mixing ratio.
In Figure~\ref{fig:couette_N2-He}, the translational temperature is displayed for the different mixing ratios.
In general, the ESBGK results using collision integrals are in excellent agreement with the DSMC results with a maximum error of $3.6\:\%$, while larger deviations up to $8.5\:\%$ are visible using Wilke's mixing rules.
This is probably due to the large mass ratio which leads to a poorer approximation of the Prandtl number.
The results slightly differ with the mixing ratio of $\mathrm{N}_2$ and $\mathrm{He}$.
In Figure~\ref{fig:couette_N2-He-rot}, the rotational temperature is compared for the different mixing ratios.
The results are similar to the results of the translational temperature due to the thermal equilibrium in the Couette flow.
Therefore, only the translational temperature plots are shown for the following test cases.

\begin{figure}
	\centering
	\includegraphics{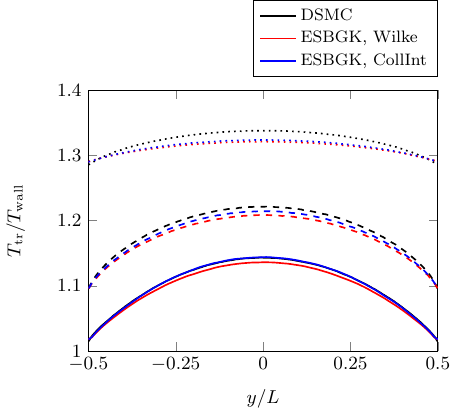}
	\caption{Comparison of the stationary solution for the temperature in a supersonic Couette flow for a $50\:\%$-$50\:\%$ $\mathrm{N}_2$-$\mathrm{N}$ mixture with $Kn_\mathrm{VHS} \approx 0.0121$ indicated with solid lines, $Kn_\mathrm{VHS} \approx 0.121$ with dashed lines and $Kn_\mathrm{VHS} \approx 1.21$ with dotted lines.}
	\label{fig:couette_N2-N}
\end{figure}

The second test mixture is a $50\:\%$-$50\:\%$ $\mathrm{N}_2$-$\mathrm{N}$ mixture, initialized with the same particle density as for the first case, $n_0=\SI{1.3e20}{\per\cubic\meter}$, as well as lower densities by one and two orders of magnitude, respectively.
This mixture was chosen because it represents a typical mass difference of species in air mixtures ($\mathrm{N}_2$,$\mathrm{N}$,$\mathrm{O}_2$,$\mathrm{O}$, $\mathrm{NO}$).
The initial densities correspond to Knudsen numbers of $Kn_\mathrm{VHS} \approx 0.0121 - 1.21$, whereby the implemented ESBGK method can be verified across different flow regimes.
The results of these simulations for the translational temperatures are shown in Figure~\ref{fig:couette_N2-N}.
Again, the results of the ESBGK simulations using collision integrals are in very good agreement with the DSMC simulations.
Due to the lower mass ratios, also Wilke's mixing rules achieve good agreement with the DSMC results.
For higher Knudsen numbers, the errors for both ESBGK methods are increasing which is also consistent with observations for the one species ESBGK model~\cite{piclas-bgk}.
Here, deviations of up to $32\:\%$ occur.
The performance of the ESBGK method thus decreases for increasing Knudsen numbers.

\begin{figure}
	\centering
	\includegraphics{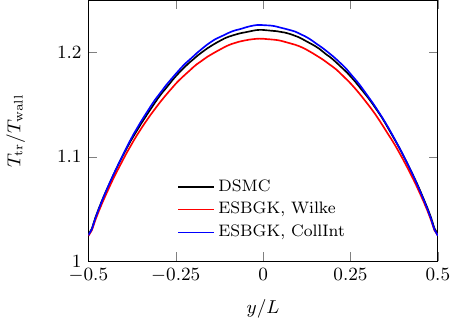}
	\caption{Comparison of the stationary solution for the temperature in a supersonic Couette flow for a $50\:\%$-$50\:\%$ $\mathrm{CO}_2$-$\mathrm{N}_2$ mixture.}
	\label{fig:couette_CO2-N2}
\end{figure}

As a third test case, a mixture with two molecular species (polyatomic and diatomic) is chosen to examine the behavior of the model with two molecular species: $50\:\%$-$50\:\%$ $\mathrm{CO}_2$-$\mathrm{N}_2$.
Again, $n_0=\SI{1.3e20}{\per\cubic\meter}$ is the initial particle density.
The results for the translational temperature are depicted in Figure~\ref{fig:couette_CO2-N2}.
All simulation methods show very good agreement with errors below $3.8\:\%$.

With test cases four and five, two more realistic atmospheric gas mixtures with a larger number of constituents are simulated to test the behavior of the model with multi-species mixtures.
From now on, only the DSMC and ESBGK solutions using collision integrals for the calculation of the transport coefficients are compared since the results using Wilke's mixing rules show larger errors, in particular when considering mixtures with large mass ratios.

\begin{figure}
	\centering
	\includegraphics{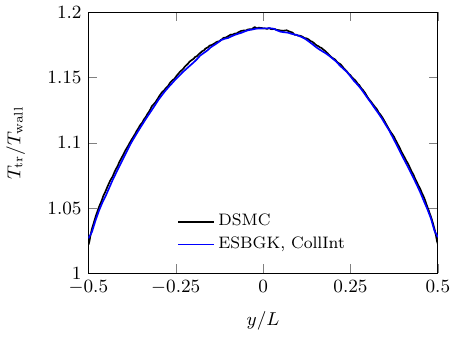}
	\caption{Comparison of the stationary solution for the temperature in a supersonic Couette flow for a $\mathrm{N}_2$-$\mathrm{O}_2$-$\mathrm{Ar}$ mixture.}
	\label{fig:couette_3spec}
\end{figure}

Case four shows the results for a $78\:\%$-$21\:\%$-$1\:\%$ $\mathrm{N}_2$-$\mathrm{O}_2$-$\mathrm{Ar}$ mixture which corresponds to a typical air composition close to the ground, again initialized with $n_0=\SI{1.0e20}{\per\cubic\meter}$.
In Figure~\ref{fig:couette_3spec}, the results for the translational temperature are compared, showing excellent agreement with deviations below $1.0\:\%$.

\begin{figure}
	\centering
	\includegraphics{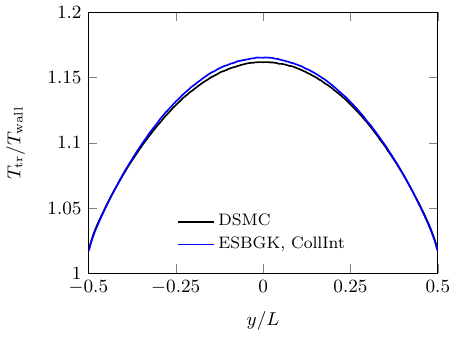}
	\caption{Comparison of the stationary solution for the temperature in a supersonic Couette flow for a $\mathrm{N}$-$\mathrm{O}$-$\mathrm{N}_2$-$\mathrm{O}_2$-$\mathrm{NO}$ mixture.}
	\label{fig:couette_5spec}
\end{figure}

For test case five, a $\mathrm{N}$-$\mathrm{O}$-$\mathrm{N}_2$-$\mathrm{O}_2$-$\mathrm{NO}$ mixture is chosen with each species contributing $20\:\%$ of the total gas with an initial particle density of $n_0=\SI{1.25e20}{\per\cubic\meter}$.
This is intended to mimic a gas composition after dissociation processes in the air.
The ESBGK results show very good agreement with the DSMC results, both illustrated in Figure~\ref{fig:couette_5spec}, with maximum deviations of $2.6\:\%$.
Both case 4 and case 5 emphasize that good results can be achieved with the model even with multi-species mixtures.

\subsection{70 degree blunted cone}
\begin{figure}
	\centering
	\includegraphics[width=0.75\linewidth]{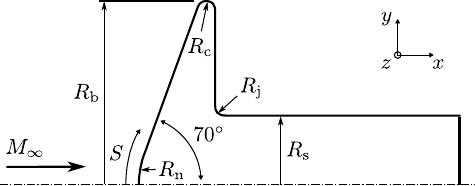}
	\caption{Geometry of the $70^\circ$ blunted cone. $R_{\mathrm{b}}=\SI{25.0}{\milli\meter}$, $R_{\mathrm{c}}=\SI{1.25}{\milli\meter}$, $R_{\mathrm{j}}=\SI{2.08}{\milli\meter}$, $R_{\mathrm{n}}=\SI{12.5}{\milli\meter}$, $R_{\mathrm{s}}=\SI{6.25}{\milli\meter}$. $S$ denotes the arc length along the surface.}
	\label{fig:70cone_geometry}
\end{figure}

To test the behavior of the ESBGK mixture model in particular in the presence of strong gradients such as across shock waves, simulations of a hypersonic flow around a 70° blunted cone are done.
The geometry of this cone is illustrated in Figure~\ref{fig:70cone_geometry}.
The simulations are performed two-dimensional axisymmetrical.
At the surface of the cone diffuse reflection and complete thermal accommodation at a constant wall temperature of $T_\mathrm{wall}=\SI{300}{K}$ is assumed.
The inflow conditions are listed in Table~\ref{t:70cone_inflow}.
\begin{table}
	\centering
	\small
	\renewcommand{\arraystretch}{1.3}
	\begin{tabular}{l|l|l|l|l}
		& Case 1 & Case 2 & Case 3 & Case 4 \\
		\hline
		Gas species & $\mathrm{N}_2$-$\mathrm{O}_2$-$\mathrm{NO}$ & $\mathrm{N}_2$-$\mathrm{O}_2$-$\mathrm{NO}$ & $\mathrm{N}_2$-$\mathrm{O}_2$-$\mathrm{NO}$ & $\mathrm{CO}_2$-$\mathrm{N}_2$ \\
		$T_\infty$ / $\si{\kelvin}$ & $13.3$ & $13.3$ & $13.3$ & $13.3$ \\
		$u_\infty$ / $\si{\meter\per\second}$ & $1502.57$ & $1502.57$ & $1502.57$ & $1502.57$ \\
		$M_\infty$ & $20.7$ & $20.7$ & $20.7$ & $23.2$ \\
		$n_\infty$ / $\si{\per\cubic\meter}$ & $3.7\cdot 10^{20}$ & $1.48\cdot 10^{21}$ & $5.92\cdot 10^{21}$ & $3.7\cdot 10^{20}$ \\
		$Kn_{\mathrm{VHS},\infty}$ & $0.147$ & $0.0366$ & $0.0092$ & $0.108$
	\end{tabular}
	\caption{Inflow conditions of hypersonic flow around a 70° blunted cone.}
	\label{t:70cone_inflow}
\end{table}
A $50\:\%$-$25\:\%$-$25\:\%$ $\mathrm{N}_2$-$\mathrm{O}_2$-$\mathrm{NO}$ gas mixture is chosen for Cases 1-3, while a $50\:\%$-$50\:\%$ $\mathrm{CO}_2$-$\mathrm{N}_2$ mixture is used for Case 4.
The first three cases were chosen to test the behavior of the model for different Knudsen numbers on the more complex case and the fourth case was selected to test the behavior with a polyatomic molecule.
The mesh contains $5064$ cells for Cases 1 and 4 with $1.06\cdot10^{6}$ total simulation particles and $16320$ cells for Cases 2 and 3 with larger particle densities in the free-stream.
For Case 2, $3.13\cdot10^{6}$ particles are in the simulation, while for Case 3, there are $7.38\cdot10^{6}$ ESBGK simulation particles and $1.45\cdot10^{7}$ DSMC simulation particles, respectively.
For the DSMC method, a time step of $\SI{1e-8}{\second}$ is used for Cases 1, 2 and 4, and a smaller time step is chosen for Case 3 with $\SI{2e-9}{\second}$, while for the ESBGK method, the time step is $\SI{2.5e-8}{\second}$.
The results using collision integrals directly and Wilke’s mixing rules are almost identical for all simulations due to the small mass differences between the gas constituents and thus good approximation with Wilke’s mixing rules, which is why the latter are not shown separately.

\subsubsection{Case 1}
For the first test case, a particle density in the free-stream of $n_\infty = \SI{3.7e20}{\per\cubic\meter}$ is assumed, corresponding to $Kn_{\mathrm{VHS},\infty} = 0.147$, to show the abilities of the ESBGK model for non-equilibrium flows.

\begin{figure}
	\centering
	\includegraphics{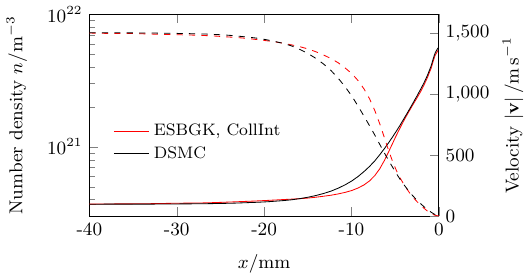}
	\caption{70° blunted cone, Case 1: Mixture mean values of velocity in $x$ direction (dashed) and number density (solid) along stagnation stream line using DSMC and ESBGK.}
	\label{fig:70degcone-lowdens-1}
\end{figure}

The simulation results of the mean flow variables of the mixture are shown in Figure~\ref{fig:70degcone-lowdens-1}.
Overall good agreement is visible between the results of both methods.

\begin{figure*}
	\centering
	\subfloat[Mean mixture temperatures along the stagnation stream line]{\includegraphics{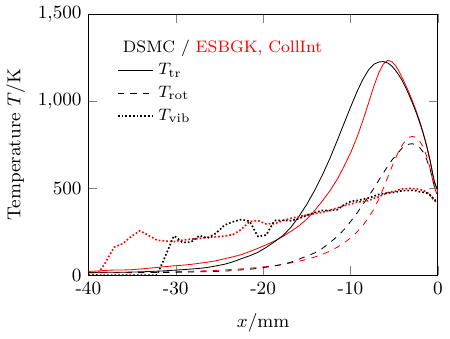}}
	\subfloat[Vibrational temperatures per species at $y=\SI{20}{\milli\meter}$]{\includegraphics{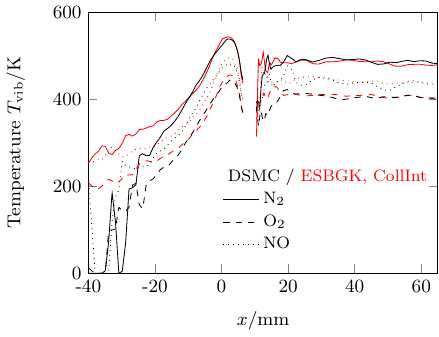}}
	\caption{70° blunted cone, Case 1: Mixture mean values of the translation, rotational and vibrational temperatures along the stagnation stream line and species-specific values of the vibrational temperature at $y=\SI{20}{\milli\meter}$ using DSMC and ESBGK.}
	\label{fig:70degcone-lowdens-2}
\end{figure*}

\begin{figure*}
	\centering
	\includegraphics{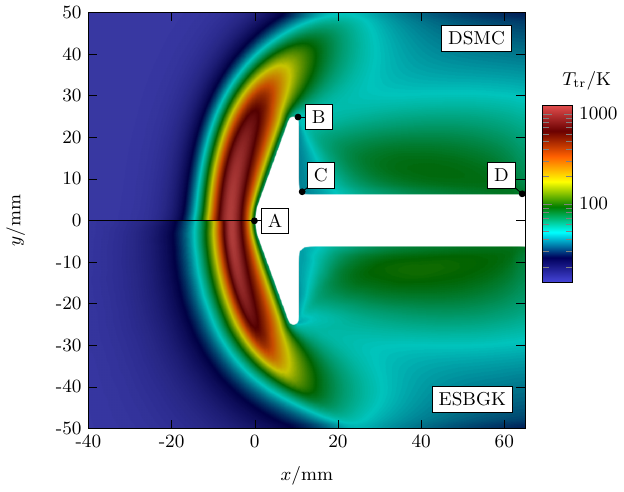}
	\caption{70° blunted cone, Case 1: Comparison of the mixture mean value of the translational temperature in the flow field using DSMC and ESBGK. Characteristic points A-D on the cone surface are indicated.}
	\label{fig:70degcone-lowdens-domain}
\end{figure*}

In Figure~\ref{fig:70degcone-lowdens-2}, the mean translational, rotational and vibrational temperatures of the gas mixture are compared for the DSMC and ESBGK simulations.
The ESBGK model predicts a earlier onset of the translational temperature increase compared to DSMC, which results in a wider shock region, also visible in Figure~\ref{fig:70degcone-lowdens-domain}, where the translational temperature in the flow field is shown in comparison.
However, this behavior is typical for the ESBGK method, even for just one species~\cite{piclas-bgk, fei2020benchmark,bgk-multispecies}.
The agreement in the post-shock region is excellent.
Figure~\ref{fig:70degcone-lowdens-2} also shows the vibrational temperatures of each species at $y=\SI{20}{\milli\meter}$.
Even in the non-equilibrium region in the wake of the 70° cone, the agreement between the results of the ESBGK and DSMC methods is very good.
Furthermore, a relatively large deviation in the vibrational temperature is visible in the front of the shock in Figure~\ref{fig:70degcone-lowdens-2}.
The reason for this is stochastic noise.
Due to the very low inflow temperature in order to achieve the high Mach number and due to the quantized treatment of the vibration, only very few particles are vibrational excited in this area, which leads to this high statistical noise.
However, using significantly more particles only makes limited sense here, as this region in terms of vibrational excitation has no influence on the behavior of the flow.
Comparing the vibrational temperatures is only meaningful closer to the shock, above temperatures of around 400~K, as a sufficient number of particles is excited here.

\begin{figure*}
	\centering
	\subfloat{\includegraphics{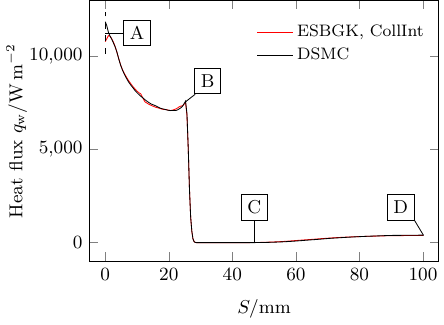}}
	\subfloat{\includegraphics{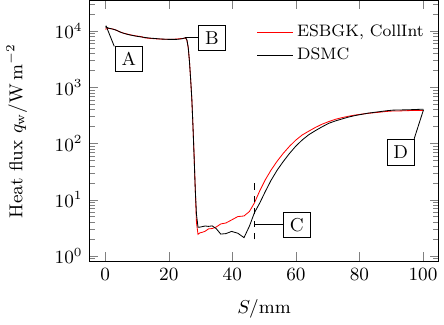}}
	\caption{70° blunted cone, Case 1: Heatflux along the surface with characteristic points A-D indicated (see Figure~\ref{fig:70degcone-lowdens-domain}).}
	\label{fig:70degcone-lowdens-3}
\end{figure*}

The heat flux on the cone surface is compared in Figure~\ref{fig:70degcone-lowdens-3}.
For both the flow-facing cone surface as well as the rear cone part, excellent agreement between both methods is shown.

\subsubsection{Case 2}
As a second test case, a particle density in the free-stream of $n_\infty = \SI{1.48e21}{\per\cubic\meter}$ is assumed, corresponding to $Kn_{\mathrm{VHS},\infty} = 0.0366$, and representing a transition regime case.

\begin{figure}
	\centering
	\includegraphics{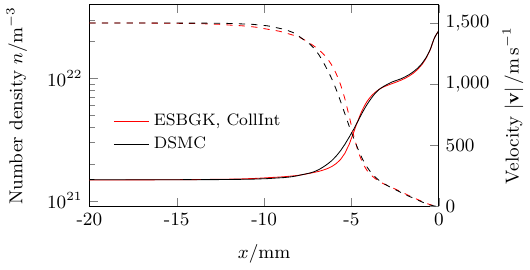}
	\caption{70° blunted cone, Case 2: Mixture mean values of velocity in $x$ direction (dashed) and number density (solid) along stagnation stream line using DSMC and ESBGK.}
	\label{fig:70degcone-middledens-1}
\end{figure}

The simulation results of the mean values of the velocity in $x$ direction and the particle density of the mixture are shown in Figure~\ref{fig:70degcone-middledens-1} with overall good agreement between the results of both methods.

\begin{figure*}
	\centering
	\subfloat[Mean mixture temperatures along the stagnation stream line]{\includegraphics{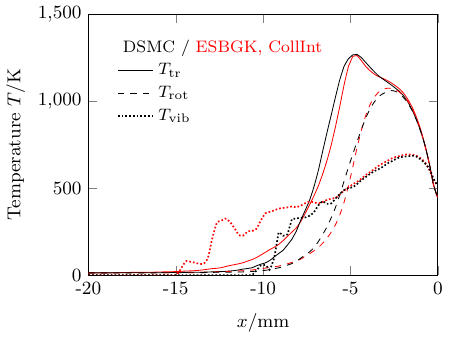}}
	\subfloat[Vibrational temperatures per species at $y=\SI{20}{\milli\meter}$]{\includegraphics{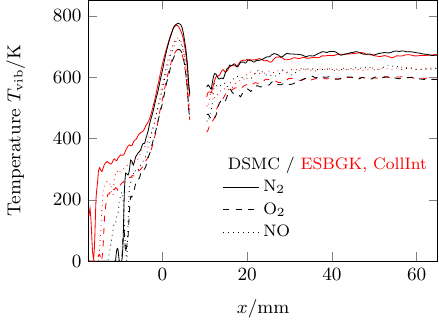}}
	\caption{70° blunted cone, Case 2: Mixture mean values of the translation, rotational and vibrational temperatures along the stagnation stream line and species-specific values of the vibrational temperature at $y=\SI{20}{\milli\meter}$ using DSMC and ESBGK.}
	\label{fig:70degcone-middledens-2}
\end{figure*}

In Figure~\ref{fig:70degcone-middledens-2}, the mean translational, rotational and vibrational temperatures of the gas mixture are compared for the DSMC and ESBGK simulations.
Similar to Case~1, the ESBGK model predicts an earlier onset of the translational temperature increase compared to DSMC, resulting in a wider shock region.
Excellent agreement is visible in the post-shock region.
Figure~\ref{fig:70degcone-middledens-2} shows the vibrational temperatures of each species at $y=\SI{20}{\milli\meter}$ additionally.
The noise behavior is also visible here, similar to Case 1.
Again, there is very good agreement between the results of the ESBGK and DSMC methods even in the non-equilibrium region in the wake of the 70° cone.

\begin{figure*}
	\centering
	\subfloat{\includegraphics{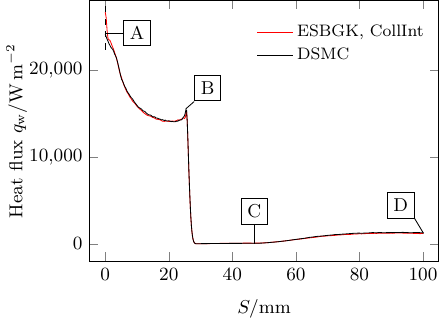}}
	\subfloat{\includegraphics{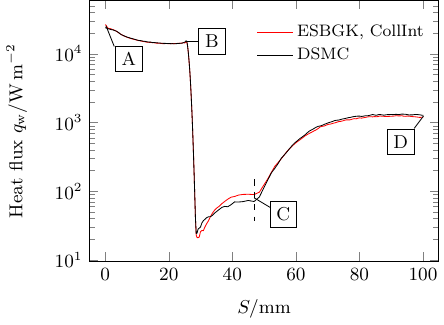}}
	\caption{70° blunted cone, Case 2: Heatflux along the surface with characteristic points A-D indicated (see Figure~\ref{fig:70degcone-lowdens-domain}).}
	\label{fig:70degcone-middledens-3}
\end{figure*}

A comparison of the heat flux on the cone surface is shown in Figure~\ref{fig:70degcone-middledens-3}.
For both the flow-facing cone surface as well as the rear cone part, both the ESBGK and the DSMC simulation results are in excellent agreement.

\subsubsection{Case 3}
The third test case represents a continuum case with a particle density in the free-stream of $n_\infty = \SI{5.92e21}{\per\cubic\meter}$, corresponding to $Kn_{\mathrm{VHS},\infty} = 0.0092$.

\begin{figure}
	\centering
	\includegraphics{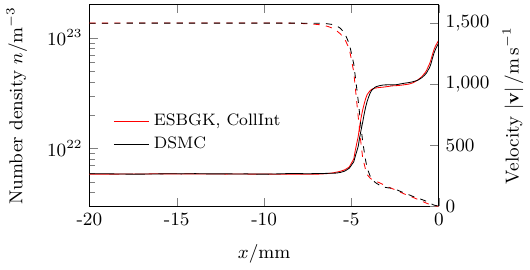}
	\caption{70° blunted cone, Case 3: Mixture mean values of velocity in $x$ direction (dashed) and number density (solid) along stagnation stream line using DSMC and ESBGK.}
	\label{fig:70degcone-highdens-1}
\end{figure}

The the mean values of the velocity in $x$ direction and the particle density of the mixture are compared in Figure~\ref{fig:70degcone-highdens-1} for the ESBGK and the DSMC models, both showing excellent agreement.

\begin{figure*}
	\centering
	\subfloat[Mean mixture temperatures along the stagnation stream line]{\includegraphics{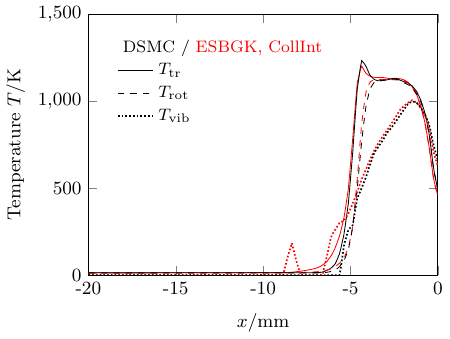}}
	\subfloat[Vibrational temperatures per species at $y=\SI{20}{\milli\meter}$]{\includegraphics{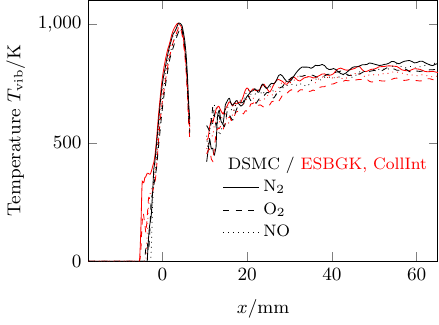}}
	\caption{70° blunted cone, Case 3: Mixture mean values of the translation, rotational and vibrational temperatures along the stagnation stream line and species-specific values of the vibrational temperature at $y=\SI{20}{\milli\meter}$ using DSMC and ESBGK.}
	\label{fig:70degcone-highdens-2}
\end{figure*}

In Figure~\ref{fig:70degcone-highdens-2}, the mean translational, rotational and vibrational temperatures of the gas mixture are shown.
The earlier onset of the translational temperature increase of the ESBGK model is again visible compared to DSMC, but the difference in the translational temperature profile is smaller compared to Cases 1 and 2.
This is due to the higher density, leading to a narrower shock region in general.
Again, excellent agreement is visible in the post-shock region.
Furthermore, Figure~\ref{fig:70degcone-highdens-2} illustrates the vibrational temperatures of each species at $y=\SI{20}{\milli\meter}$, showing good agreement with only little deviations in the non-equilibrium region in the wake of the 70° cone.

\begin{figure*}
	\centering
	\subfloat{\includegraphics{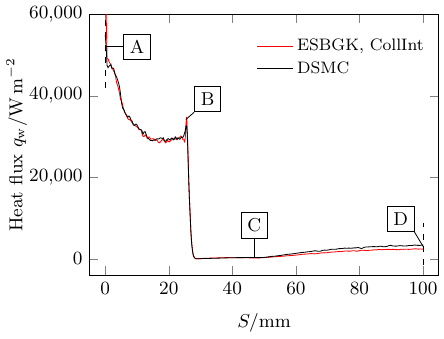}}
	\subfloat{\includegraphics{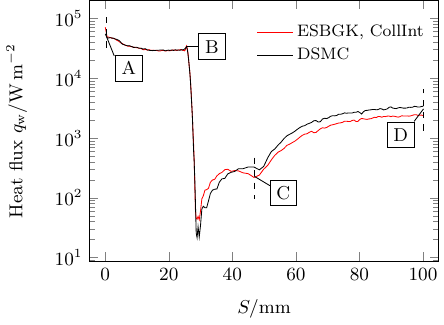}}
	\caption{70° blunted cone, Case 3: Heatflux along the surface with characteristic points A-D indicated (see Figure~\ref{fig:70degcone-lowdens-domain}).}
	\label{fig:70degcone-highdens-3}
\end{figure*}

The simulation results for the heat flux on the cone surface is compared in Figure~\ref{fig:70degcone-highdens-3} for both of the models.
There is good agreement of the results for the flow-facing cone surface as well as only little deviations for the rear cone part.

\subsubsection{Case 4}
For Case 4, a $50\:\%$-$50\:\%$ $\mathrm{CO}_2$-$\mathrm{N}_2$ is chosen to evaluate the accuracy of the ESBGK model for a polyatomic-diatomic mixture with a higher number of degrees of internal freedom in particular in vibration.
A particle density in the free-stream of $n_\infty = \SI{3.7e20}{\per\cubic\meter}$ is chosen similarly to Case 1, which corresponds to $Kn_{\mathrm{VHS},\infty} = 0.108$ for this mixture.

\begin{figure}
	\centering
	\includegraphics{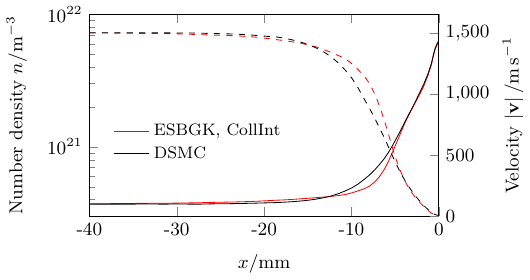}
	\caption{70° blunted cone, Case 4: Mixture mean values of velocity in $x$ direction (dashed) and number density (solid) along stagnation stream line using DSMC and ESBGK.}
	\label{fig:70degcone-CO2-N2-1}
\end{figure}

In Figure~\ref{fig:70degcone-CO2-N2-1}, the simulation results of the mean values of the velocity in $x$ direction and the particle density of the mixture are compared for the DSMC and ESBGK models.
Same as for the $50\:\%$-$25\:\%$-$25\:\%$ $\mathrm{N}_2$-$\mathrm{O}_2$-$\mathrm{NO}$ gas mixture from Case 1, there is overall good agreement.

\begin{figure*}
	\centering
	\subfloat[Mean mixture temperatures along the stagnation stream line]{\includegraphics{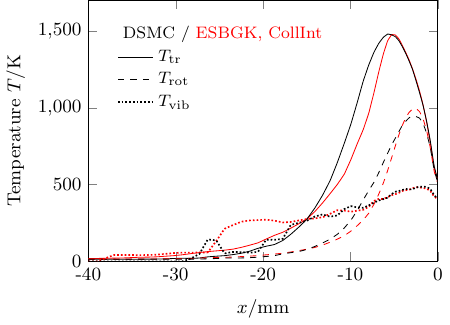}}
	\subfloat[Vibrational temperatures per species at $y=\SI{20}{\milli\meter}$]{\includegraphics{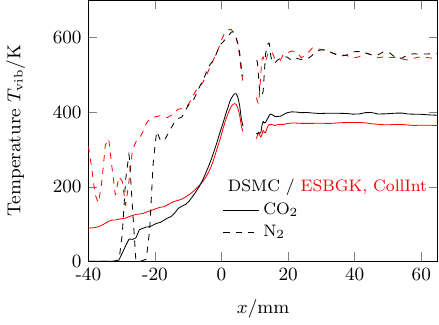}}
	\caption{70° blunted cone, Case 4: Mixture mean values of the translation, rotational and vibrational temperatures along the stagnation stream line and species-specific values of the vibrational temperature at $y=\SI{20}{\milli\meter}$ using DSMC and ESBGK.}
	\label{fig:70degcone-CO2-N2-2}
\end{figure*}

The mean translational, rotational and vibrational temperatures of the gas mixture are illustrated in Figure~\ref{fig:70degcone-CO2-N2-2}.
Again, a wider shock region is visible for the ESBGK results due to an earlier onset of the translational temperature increase compared to DSMC, while there is very good agreement in the post-shock region.
Also, the vibrational temperatures of each species at $y=\SI{20}{\milli\meter}$ are shown.
While there is excellent agreement between the results of the ESBGK and DSMC methods in the non-equilibrium region in the wake of the 70° cone for $\mathrm{N}_2$, a small deviation can be seen for $\mathrm{CO}_2$.
It is difficult to pinpoint the reason for this.
Perhaps the behavior of the polyatomic vibratory excitation with the very large Knudsen numbers in the wake can simply no longer be reproduced with the model.

\begin{figure*}
	\centering
	\subfloat{\includegraphics{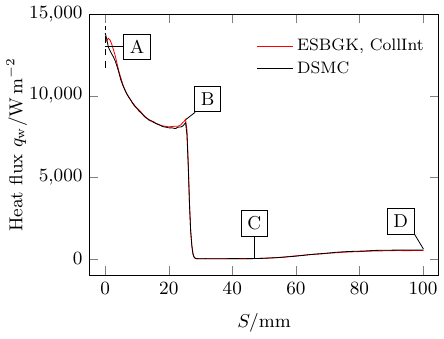}}
	\subfloat{\includegraphics{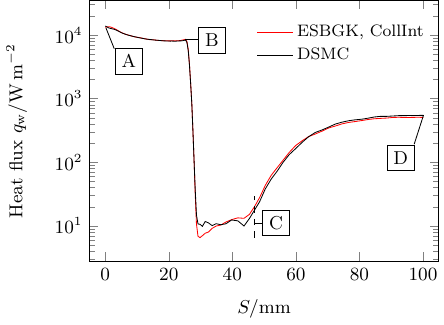}}
	\caption{70° blunted cone, Case 4: Heatflux along the surface with characteristic points A-D indicated (see Figure~\ref{fig:70degcone-lowdens-domain}).}
	\label{fig:70degcone-CO2-N2-3}
\end{figure*}

The comparison of the heat flux on the cone surface in Figure~\ref{fig:70degcone-CO2-N2-3} shows very good agreement.

\subsection{Computational performance}
It is expected that the resolution requirements for the ESBGK method are less restrictive than for the DSMC method, depending on the Knudsen number, which allows for computational time savings as briefly described in the introduction.
Thus, the basic idea is to apply the ESBGK method in regions of low Knudsen numbers to gain computational efficiency compared to the DSMC method.
The aim is then to use this particle-based continuum method for a coupling with the DSMC method in order to solve multi-scale problems.
In the following, the resolution requirements of both of the methods are discussed in greater detail.

The spatial resolution of the BGK particle method is essentially determined by the gradients in the flow and thus by the decoupling of the particle movement and the collision operator.
However, this is not necessarily coupled to the mean free path as in the case of DSMC, which is why in denser cases, significantly fewer particles can be used in a simulation.
To achieve an optimal spatial resolution for the BGK method, appropriately adapted grids would be needed, as described in~\citet{fp}.
However, all simulations for both the ESBGK and the DSMC methods presented in this paper were performed on the same grid to be able to test the physical accuracy of the ESBGK model.
Therefore, the computational optimum of both methods could not be achieved here.
Nevertheless, regarding for example the simulations of Case~3, half as many particles could be used in the ESBGK case compared to DSMC, leading to a higher computational efficiency.
This difference in the particle number would become even more pronounced in a 3D simulation case, since the particle number required to resolve the mean free path in the DSMC method increases with the power of the dimension, while the BGK method instead only needs to resolve the gradients in flow quantities.

Regarding temporal resolution, the mean collision time needs to be resolved for the DSMC method while for BGK, the stiff relaxation frequency needs to be resolved instead, at least when an explicit first-order time discretization is used.
A few methods exist for particle BGK methods to circumvent this resolution criterion~\cite{PhysRevE.106.025303,FEI2020108972,fei2021efficient}.
Nevertheless, in the simulations here, a simple explicit time integration was used.
However, the collision frequency in the DSMC method and the relaxation frequency in the BGK method differ.
As described in~\citet{MATHIAUD202265}, both vary depending on the collision model used.
The relaxation frequency of the ESBGK model scales with the Prandtl number, often allowing for larger time steps even with explicit time integrations, which in the end leads to a higher computational efficiency.

Considering all the above-mentioned arguments, one concludes that the ESBGK method should become computationally more efficient than the DSMC method when solving smaller Knudsen numbers flows.
This is also reflected in the simulations presented above, where in Case~1, the computational time could be reduced by a factor of 10 with the ESBGK method compared to the DSMC method, while in Case~3, due to the smaller Knudsen number, a factor of 28 could be saved.
However, for larger Knudsen numbers, the DSMC method will be faster than the ESBGK method at some point, because only few or even no collisions need to be computed anymore.
A more detailed comparison of computational times goes beyond the scope of this paper, as it mainly focuses on the theoretical foundations and physical results of the model.

\section{Conclusion}
An ellipsoidal statistical BGK mixture method is proposed and implemented in the open-source particle code PICLas based on the model of \citet{MATHIAUD202265,energy-cons} and \citet{brull, brull2021ellipsoidal}.
Multi-species gas flows are simulated while allowing for the treatment of internal energies of di- and polyatomic molecules, aiming for a coupling with the DSMC method to solve multi-scale problems.
As test cases for verification, different supersonic Couette flows and flows around a 70° blunted cone are used, where in general, overall good agreement between the ESBGK and DSMC methods is achieved for different type of mixtures and Knudsen numbers.
Deviations occur concerning the onset of the temperature increase in the shock region in front of the 70° blunted cone, which is a known behavior for the ESBGK method.
Visible deviations in the translational temperature profile in the shock region become smaller looking at smaller Knudsen number flows with narrower shock regions.
The agreement of the simulation results in both the post-shock region and on the cone surface i.e. the heat flux are excellent.
Minor deviations are visible in the wake of the 70° blunted cone for the vibrational temperature of $\mathrm{CO}_2$ which is expected to be related to the polyatomic vibratory excitation in this high-Knudsen regime.
Additionally, two methods for the determination of the transport coefficients of a gas mixture, Wilke's mixing rules and collision integrals, are compared, whereby the latter shows larger errors for mixtures with large mass ratios.
A considerable reduction of the computational duration in transition and continuum regimes is possible with the ESBGK method compared to DSMC due to less restricted resolution requirements of the BGK method.

In future work, the implementation of chemical reactions into the model is envisioned.
Furthermore, an extension of the already implemented atomic Shakhov model to polyatomic gas mixtures will be done.

\section*{Acknowledgments}
This project has received funding from the European Research Council (ERC) under the European Union’s Horizon 2020 research and innovation programme (grant agreement No. 899981 MEDUSA).



\bibliographystyle{elsarticle-num-names} 
\bibliography{mybibfile}


\end{document}